\documentclass[preprint,12pt]{elsarticle}

\usepackage{amssymb}

\usepackage{amsmath}

\usepackage{standalone}
\standaloneconfig{mode=buildnew}
\usepackage{amsthm,amsfonts}
\usepackage{bm}
\usepackage{subfig}
\usepackage{multirow}
\usepackage{makecell,tabularx}
\usepackage[utf8]{inputenc}
\usepackage{morefloats}
\usepackage{xr-hyper}
\usepackage[normalem]{ulem}
\usepackage{xcolor}

\usepackage{graphicx}

\newcommand{\iid}{\stackrel{\rm iid}{\sim}}
\newcommand{\ind}{\stackrel{\rm ind}{\sim}}

\newcommand{\e}{\mathrm{e}}

\DeclareMathOperator{\diag}{diag}

\graphicspath{{Figures/}}
\journal{Computational Statistics and Data Analysis}

\begin{document}
	
	\begin{frontmatter}
		
		
		
		\title{A Bayesian time-varying random partition model for large spatio-temporal datasets} 
		

		\author[1]{Giulio Beltramin} 
		\author[2]{Andrea Cremaschi} 
		\author[3]{Annalisa Cadonna} 
		\author[1]{Alessandra Guglielmi} 
		\author[4]{Fernando Andrés Quintana}

		\affiliation[1]{organization={Department of Mathematics, Politecnico di Milano},
			city={Milan},
			country={Italy}}
		\affiliation[2]{organization={School of Science and Technology, IE University},
			city={Madrid},
			country={Spain}}
		\affiliation[3]{organization={Department of Statistics, University of Klagenfurt},
		city={Klagenfurt},
		country={Austria}}
		
		\affiliation[4]{organization={Departamento de Estadística, Facultad de Matemáticas, Pontificia Universidad Católica de Chile},
			city={Santiago},
			country={Chile}}

		\begin{abstract}
	Spatio-temporal areal data can be seen as a collection of time series which are spatially correlated, according to a specific neighbouring structure. Motivated by a dataset on mobile phone usage in the Metropolitan area of Milan, Italy, we propose a semi-parametric hierarchical Bayesian model allowing for time-varying as well as spatial model-based clustering.
	Our approach incorporates the notion of regimes that describe changing patterns over work and night hours as well as
	weekdays/weekends. Changes across regimes are considered by means of temporal changepoint components that allow for different hierarchical structures specified across time points. The changepoints might occur within fixed time windows over the
	day. The model features a novel random partition prior that incorporates the desired spatial features and encourages co-clustering based on areal proximity. We explore properties of the model by way of extensive simulation studies from which we collect valuable information. Finally, we discuss the application to the motivating data, where the main goal is to spatially cluster population patterns of mobile phone usage.  
		\end{abstract}
		
		
		
		
	\end{frontmatter}
	
		

\section{Introduction}
\label{sec:intro}

In recent years, there has been a rapid increase in the availability of data collected over time on areal units. Areal units can be defined by geographical boundaries (e.g., regions, counties, municipalities), or by tessellating the territory of interest. The available data are often large-to-huge collections of (possibly long) time series that are spatially correlated, according to a specific neighbourhood structure. Data of this type are extremely helpful to understand population distribution in an urban space, which is critical for urban planning and provision of municipal services; see, e.g. \cite{deville2014dynamic}, \cite{sulis2018using}, and \cite{tu2020portraying}. However, traditional methods for
exploring human population dynamics, such as censuses and surveys, can be very expensive. Moreover, more detailed changes over time, such as daily commute and urban transportation, are challenging to assess with traditional methods, but crucial nonetheless, to urban planning. 

Mobile phone-based data have been extensively used to extract geographical knowledge in previous studies; see, for instance, \cite{song2010limits} and \cite{liu2015social}. Previous studies \citep[see][and the references therein]{wang2021bayesian} suggest that mobile phone data are a valid proxy for human activities and interactions. 

Motivated by the study of population density dynamics arising in the context of area-level mobile data, we specify an appropriate model and develop efficient algorithms for the statistical analysis of large-to-huge spatio-temporal areal data. The proposed hierarchical model takes into account various characteristics of the data, including (a) varying regimes corresponding to day/night and weekday/weekend times; (b) spatial dependence across areas; and (c) clustering of areal time series. We accomplish these goals by employing a novel Bayesian nonparametric prior for clustering longitudinal areal data.

The motivating dataset has been previously considered in \cite{Manfredini2015} and \cite{Secchi2015}, where each time series is modeled as the realization of a functional data process. One of the main goals of this work is to model the population dynamics, and in particular, to understand how this changes across different regimes, and thus our approach does not employ a functional data perspective but rather builds on a combination of harmonic regression models for the temporal component and conditionally autoregressive (CAR) models for the spatial association. CAR models are commonly used to represent spatial autocorrelation for areal data, and can be thought of as the conditional specification of a (Gaussian) Markov random field (GMRF). A comprehensive review of GMRF models can be found in \cite{Rue:2005:GMR:1051482}. CAR models were originally introduced as spatial models in \cite{besag1974spatial, Besag1975}, and they have been used since as the likelihood for the observations themselves in one-stage models, or as the distribution of spatial random effects, as part of a Bayesian hierarchical model \citep[see][]{Besag1991}. The different CAR
models proposed in the literature correspond to specific choices of the precision matrix for the corresponding GMRF. Thanks to the Markov property, the precision matrix of a GMRF is potentially very sparse, which enables efficient computation through linear algebra algorithms for sparse matrices. An efficient algorithm for block updating in the context of Markov Random Fields models is introduced in \cite{knorr-held2002}. Different CAR prior specifications have been proposed, such as the intrinsic and Besag-York-Molli\'e priors \citep[both in][]{Besag1991}. We adopt instead the formulation proposed in \cite{Leroux2000}, which is useful to
estimate spatial correlation among the random effects. 

The current literature on spatial clustering does not include the partition of areal units in the random parameters of the model, but obtains cluster estimates through the application of empirical clustering methods (e.g. hierarchical clustering) to posterior summaries obtained from the data. 
In particular, focusing on the literature on non-Bayesian statistical and machine learning techniques, we mention spatio-temporal K-means \citep{dorabiala2022spatiotemporal}, a 2-step procedure that creates
clusters first using a K-means procedure that considers clusters at two
consecutive time points and then classifies trajectories by assessing which
cluster appears more frequently over time. The ST-DBSCAN algorithm for spatial–temporal data is employed in \cite{birant2007st} by applying
DBSCAN to  spatial and temporal features in a simultaneous and independent
fashion. 
A hierarchical clustering method with spatial constraints, \texttt{ClustGeo} is used in \citet{chavent2018clustgeo}, based on Ward hierarchical clustering. This is a 2-step procedure that first
applies \texttt{hclust} to a feature matrix (e.g., variables of interest), 
and then ``corrects'' the clustering by taking as dissimilarity matrix a linear combination of the original dissimilarity (based on features) and the one computed with the geospatial information. Another approach based on hierarchical clustering is presented in \cite{heaton2017nonstationary}, where Gaussian processes are employed to handle nonstationary time series. The mixed geographically weighted regression with spatial auto-correlation method \texttt{MGWRSAR} is implemented in the R package \texttt{mgwrsar} \citep{mgwrsar}. The authors consider SAR-type models with various combinations of constant and spatially-varying coefficients. A combination of geographically
weighting with artificial neural networks, \texttt{GWANN}, is proposed by
\citep{hagenauer2022geo}. Neither \texttt{MGWRSAR} nor \texttt{GWANN} directly provide cluster estimates, but one can naively apply clustering techniques to the estimated spatial effects or predicted values generated from these techniques; see Supplementary Section 3.
Instead, the main advantage of our approach is that we directly model the partition of areas and make this a central object of posterior inference. Our prior, here denoted areal product partition model (aPPM), penalizes excessive areal disconnectedness
through a spatial association parameter $\xi$. The prior is new in the BNP literature for random partition models, though it builds on ideas in \cite{Hegarty_Barry2008}; see also \cite{pavani2025multivariate} for an application to multivariate analysis of data from mosquito-borne diseases in Brazil that employs the Hegarty and Barry prior. Note that product partition models for estimating change points for a single time series were originally introduced in \cite{barry1992product}. 

Finally, a different Bayesian nonparametric model for clustering has been proposed in \cite{cadonna2019bayesian} using the same dataset analyzed in this work. There are two main differences between their model and the one proposed in this work. First, in \cite{cadonna2019bayesian}, areal units are clustered together if their behaviour is similar over the entire time series under analysis, while we allow for clustering changes across regimes.
Second, \cite{cadonna2019bayesian} use an ANOVA-DDP prior which does not consider the neighbouring structure of the areal units, while the aPPM introduced in Section~\ref{sec:aPPM} explicitly takes this into account.

Our contributions to the analysis of spatio-temporal
areal datasets can be summarized as follows: (i) we propose a novel spatio-temporal random partition model that incorporates well studied priors in the Bayesian parametric and nonparametric framework and that provides a satisfactory fit to the available data; (ii) we estimate spatially-driven clusters of areal regions, and assess their changes over regimes; (iii) we provide inference on changepoints, where the dynamics of mobile phone usage transit between different regimes; (iv) we are able to handle missing data. 

The remainder of the paper is organized as follows. Section~\ref{sec:data} describes the data and motivations. Section~\ref{sec:Model} introduces the modelling framework and the proposed aPPM prior. Section~\ref{sec:application-telecom} presents the application to population density dynamics in Milan, and Section~\ref{sec:conclusions} concludes. A Supplementary Materials file is provided, containing details on the tailored MCMC algorithm, simulation studies, additional results and figures, as well as comparisons with alternative approaches.

\section{Description of the Dataset and Motivation}
\label{sec:data}

Motivated by the study of population density dynamics in the metropolitan area of Milan, Italy, we propose a model for spatio-temporal clustering of areal data. We consider data as in \cite{Secchi2015} (more details are reported below),  though we focus on a smaller grid to understand the clustering of areas in the inner part of the city, which should be easy to interpret in terms of the city's particular features such as entertainment and restaurant areas, or neighbourhoods that are mostly residential. The municipality of Milan, located in the metropolitan area centre, is a major productivity and financial centre for the entire northern portion of Italy. About 1.3 million people live and 600 thousand people commute every working day between Milan and the metropolitan area. Using mobile phone data, 
we study how the dynamics of the population density evolves over time, potentially uncovering temporal and spatial patterns. Moreover, we would like to partition the areas covered by the data into subregions sharing a similar population dynamic pattern across the observed time window. This is of great interest to urban planners, city managers and network providers. For instance, the data in \cite{Secchi2015} is also analyzed in \cite{Manfredini2015}, who investigated relevant urban usage by proposing diversified management policies for increased efficiency of public services supply. \cite{wang2021bayesian} propose a Bayesian spatio-temporal model focusing on area-level mobile phone users data. Specifically, their data are the total number of mobile phone users actively recorded by cell towers in one-hour intervals in an overall time window of 24 hours in Shenzhen, China. They do not assume a random partition, and infer the clustering structure of the spatial data via a hierarchical clustering procedure on the estimated time profiles. They nevertheless conclude that their Bayesian spatio-temporal model can enhance the understanding of the space-time variability of population distribution using mobile phone data.

The dataset we consider is provided by Telecom Italia, the largest mobile company in Italy, as part of the Green Move Initiative (financed by Regione Lombardia), through a research agreement between Telecom and Politecnico di Milano. See further details in \cite{Secchi2015}. These authors segment the metropolitan area of Milan into subregions that share the same activity pattern along time in terms of population density dynamics. To this end,  they integrate a Treelet analysis for dimensional reduction with a Bagging Voronoi
strategy for the exploration of spatial dependence, in order to reduce the dimension of spatially dependent signals. They propose the Bagging Voronoi Treelet algorithm, that decomposes the massive dataset ($10,573$ areas for $1,308$ time points, resulting in more than 13 millions records) into relevant spatial and temporal dynamics.

In this work, we focus on a portion of the metropolitan area of Milan between $45.444^{\circ}$ and $45.49^{\circ}$ in latitude and between $9.15^{\circ}$ and $9.225^{\circ}$ in longitude. This portion covers approximately the area inside a city belt called \textit{circonvallazione} (i.e., circumvallation), hosting a significant flow of private and public transportation vehicles during rush hours. The districts inside this belt line can
be considered the city centre. We have partitioned the central area into a grid of
$I=13\times 14$ sites (areal units).

\begin{figure}[ht!]
	\begin{center}
	\subfloat[]{\includegraphics[width=0.575\textwidth]{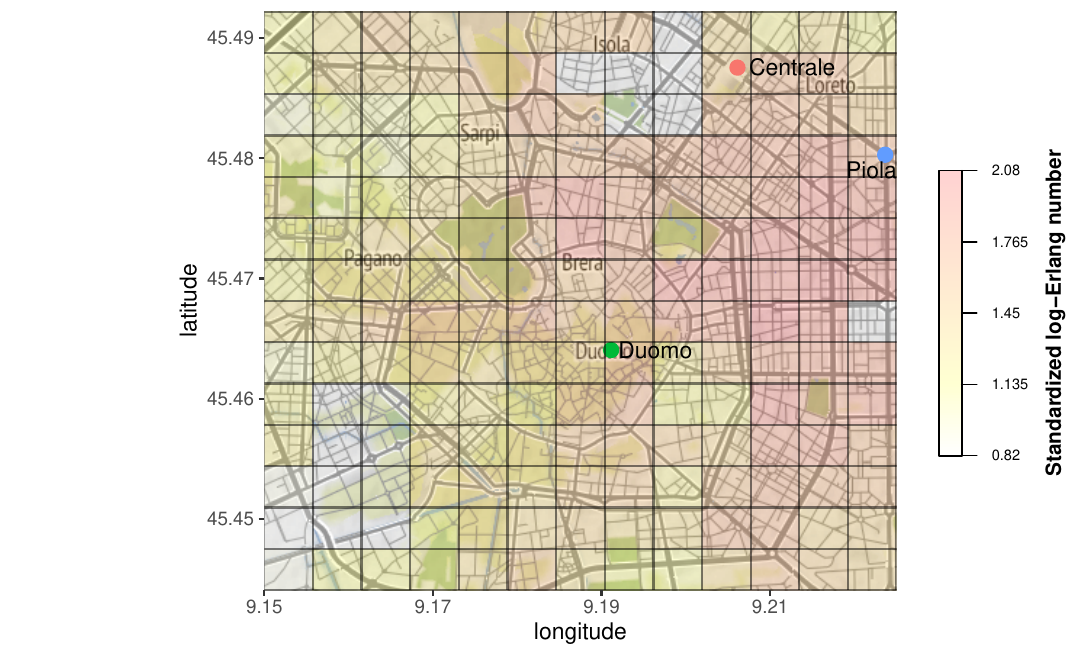}}
	\subfloat[]{\includegraphics[width=0.392\textwidth]{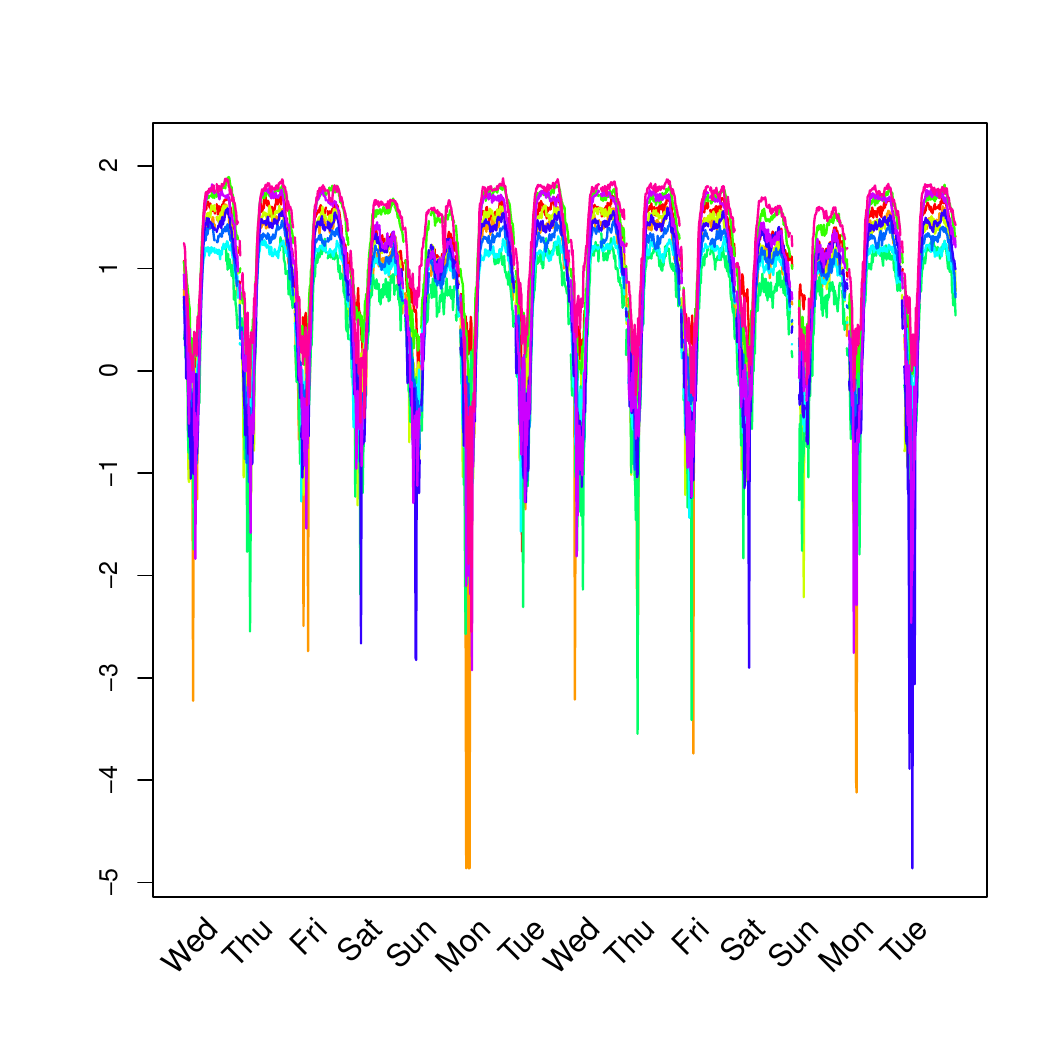}}
	\end{center}
	\caption{Telecom data. (a) Data recorded on Wednesday March 18th 2009, at noon. The gray areas correspond to missing observations. Points of interests as Piazza Duomo, Stazione Centrale and Piazza Piola are denoted as colored points on the map. (b) Time series for ten randomly selected locations. Data shown are standardised $\log$-Erlang numbers for all areal units.} 
	\label{fig:data_example}
\end{figure}

Each areal unit was recovered from the original data by Telecom, putting together 4 of the original sites from a uniform lattice of an area including the metropolitan area though we consider only inner areal units as stated above. See Figure~\ref{fig:data_example} (a) presenting the portion of the central municipality of Milan under study. The data are recorded every 15 minutes from March 18th 2009, 00:15 to March 31st 2009, 23:45, yielding a time series of length $T = 1343$ for each of the lattice sites. In total, $244,608$ records are available of which $22,068$ are missing. We analyze the Erlang number, calculated as the sum of lengths of all calls in a given time interval, divided by the length of the interval. In other words, the Erlang number is equivalent to the average number of mobile phones simultaneously calling through the network, and can be considered proportional to the number of active users \citep[see][for more details]{Secchi2015}. The
Erlang number, recorded over all areal units in a region, can be used as a proxy for population density in that region, and through its changes over time, of population density dynamics.

Figure~\ref{fig:data_example}(a) shows the standardised $\log$-Erlang number recorded on Wednesday, March 18th 2009 at noon, for the selected portion of the metropolitan area.  The financial district in the centre of the city is identifiable, as it is the area with high mobile activities during working hours, as well as the eastern area corresponding to a busy portion of the city. We can also identify, for example, peripheral areas with less mobile traffic in the western part. Locations on the grid corresponding to missing observations are left blank, showing only the underlying map of the city. In what follows, we always report summary statistics or plots of globally standardised $\log$-Erlang numbers.
 
Figure~\ref{fig:data_example}(b) shows the time series for ten randomly selected locations. As we can see, the mobile activity is higher during the day hours and lower at night. Moreover, we can see differences between workdays and weekends. We consider these temporal patterns in the proposed model in such a way that the clustering structure changes over time through a simplified approach based on the notion of temporal regimes (workdays and weekends, days and nights). Another distinguished feature of our Bayesian model is that we can handle missing responses in a natural fashion unlike, for instance, \cite{Secchi2015}.

\section{Model specification}
\label{sec:Model}

We can frame the problem considered as the clustering of the same $I$ items (the 182 areal units in Milano) at different times $t=1,\ldots,T$. Hence, the parameters of interest are the random partitions of the same total area at each time point $t=1,\ldots,T$.  We could thus use the model to estimate one partition
for each time $t$. By the spatial connotation of the data, it is clear that the marginal prior of any random partition has to incorporate information on the spatial structure.  By this we
mean that the probability that two neighbouring areas co-cluster should be larger than the probability of including in the same cluster two areas far apart. However, because $T=1343$ is a large number and we expect to see little or no variation between the cluster estimates separated by 15 minutes (the gap time between observations), we focus on the notion of regimes and assume that there are change points over time where the key parameter, the random partition of the same $I$ areal units, changes, as in changepoint models for a single time series.  

We now introduce the spatio-temporal hierarchical model for the motivating dataset, which has three main components, namely: (1) the	 likelihood part for (standardised) log-Erlang numbers, based on a mixed regression with time-varying coefficients and spatially-correlated random effects,  in Section~\ref{sec:likelihood_spatialeffects}; (2) the random partition prior specification for areal units, which incorporates information on the spatial structure for each regime in Section~\ref{sec:aPPM}; and (3) the
prior specification for the regimes (Section~\ref{sec:st_clustering}). The final dynamic Bayesian model is described in Section~\ref{sec:st_clustering}. 

\subsection{Likelihood and spatial random effects}
\label{sec:likelihood_spatialeffects} 
 
Consider $I$ areal units, in our case given by tessellations of part of the city of Milan, as described in Section~\ref{sec:data}. Denote by $Y_{it}$ the observation for areal unit $i=1, \ldots, I$ at time $t=1,\ldots, T$, yielding a dataset of $I \times T$ observations. In our application, the response variable $Y_{it}$ is the standardised logarithm of the Erlang number in area $i$ at time $t$. To avoid null Erlang numbers, before applying the logarithmic transformation to the data, we transform them by adding a quantity equal to the smallest observed non-zero Erlang number. It is important to remark that a zero Erlang number does not mean zero activity, but that this fell below a certain detection threshold. Then, the $\log$-transformed data are standardised across all areas and timestamps to have mean $0$ and standard deviation $1$. We applied these transformations to ensure that the data support is not limited to the positive reals, allowing the observations to be modeled as Gaussians. For each areal unit $i = 1, \dots, I$, and time $t = 1, \dots, T$, we model observation $Y_{it}$ as:
\begin{equation*}
Y_{it} \mid \bm x_t, \tilde{\bm \beta}_{i t}, \tilde{u}_{it} = \bm x'_t \tilde{\bm \beta}_{i t} + \tilde{u}_{it} + \epsilon_{it},  \quad
\epsilon_{it} \ind \text{N}\left(0, \sigma^2_{\epsilon_t}\right)
\end{equation*}
where $\epsilon_{it}$, for $i = 1, \dots, I$ and $t = 1, \dots, T$, are conditionally independent spatio-temporal residuals. The zero-mean $\tilde{u}_{it}$'s are defined below. Here, $\text{N}\left(\mu,\sigma^2\right)$ denotes the Gaussian distribution with mean $\mu$ and variance $\sigma^2$. The $p$-dimensional extension of it will be used later and indicated as $\text{N}_p\left(\bm \mu, \bm \Sigma\right)$ for mean vector $\bm \mu$ and covariance matrix $\bm \Sigma$ for any positive integer $p$. The variance of $\epsilon_{it}$ in the full Bayesian model will depend on the regime associated to $t$.
The mean surface in this model is given by $\bm x'_t \tilde{\bm \beta}_{it}$, through which we model, at each location, the periodicity of the data by resorting to a harmonic regression. Concretely, we let $\bm x_{t}$ denote a $p$-dimensional design vector, chosen from  the harmonic functions $\left(\cos(\omega_j t), \sin(\omega_j t)\right)$, where $\omega_j= 2 \pi j/T$ and $j=1,2,\ldots,T/2$. If $T$ is not an even number, we assume $T$ to be the next even number and the corresponding observation to be missing. See Supplementary Section 1 for details on Bayesian imputation of the missing Erlang numbers. 
The dataset of interest has $T=1344 =2\times 7\times 24\times 4$ time measurements, since our time span is two weeks with four observations per hour.  Since our responses are standardised, we exclude the intercept term in $\bm x_t$. 
Using harmonic regression corresponds to approximating each underlying
signal through a trigonometric polynomial. Motivated by the characteristics of our dataset and pragmatic knowledge of Milan vehicle traffic, we choose to select weekly, daily, semi-daily and hourly frequencies in $\bm x_{t}$. This is equivalent to assuming $p=8$, and 
\begin{align}\label{eq:x_harmonics}
\bm x_{t} = &\left(\cos(\omega_2 t), \sin(\omega_2 t), \cos(\omega_{14} t), \sin(\omega_{14} t), \right.\\
&\left. \cos(\omega_{28} t), \sin(\omega_{28} t),  \cos(\omega_{336} t), \sin(\omega_{336} t) \right) \nonumber
\end{align}
For instance, to set the daily frequency, we assume $j = \frac{1344}{24\times 4} = 14$, so that $\omega_{14}= \frac{2\pi}{96}$. In this case $\cos(\omega_{14} (t+96))=\cos(\omega_{14} t)$, i.e. the function has period equal to 96, which corresponds to one day interval (4 observations in an hour for 24 hours). The vector $\tilde{\bm \beta}_{i t} = \left(\tilde{\beta}_{i t, 1 }, \dots ,\tilde{\beta}_{it,p}\right)'$ is the vector of harmonic coefficients and, as discussed later in Section~\ref{sec:st_clustering}, will be used to cluster the areal units.

Having controlled seasonality through $\bm x_t$ in our model, we now consider spatial autocorrelation. This is done via a spatial random effects vector $\tilde{\bm u}_{t} = \left(\tilde{u}_{1t}, \ldots, \tilde{u}_{It}\right)'$ on which we put a spatial CAR prior, that is:
\begin{equation}\label{eq:sprandeff}
\tilde{\bm u}_{t}\mid \tau^2, Q\left(\zeta_t, W\right), \zeta_t \sim \text{N}_I\left(\bm 0,  \tau^2 Q(\tilde{\zeta}_t, W)^{-1}\right)
\end{equation}
where $W$ is the $I \times I$ matrix encoding the contiguity structure of the $I$ areal units specified as $W_{i,j}= 1$ if areal units $i$ and $j$ are neighbours and $W_{i,j}= 0$ otherwise. Here, we specifically define the neighbours of a site $i$ as the 8 cells surrounding $i$ in a grid layout. To help the description of $W$, we list its minor diagonals from the main diagonal to the bottom left corner. The only minor diagonals whose elements differ from zero are the first, 12th, 13th and 14th minor diagonals, while those from the 2nd to 11th contain only zero values, as do those from the 15th to 182th. Moreover, the matrix $W$ is a block-tridiagonal matrix with $14$ blocks of dimension $13\times 13$, with each block being tridiagonal itself. We next specify $Q$ in \eqref{eq:sprandeff} following the construction discussed in~\cite{Leroux2000}, that is, we set $Q\left(\tilde{\zeta}_t, W\right)= \tilde{\zeta}_t (\diag(W \bm 1) - W) + (1-\tilde{\zeta}_t) \mathbb I_I$, where $\mathbb{I}_I$ is the $I$-dimensional identity matrix and $\bm 1$ is an $I$-dimensional vector of ones. Let $d_i$ be the number of neighbours of site $i$. The matrix $(\diag(W \bm 1) - W)$ has elements equal to $d_i$ if $i=j$, equal to $-1$ if $i$ and $j$ are neighbours ($i \sim j$), and equal to $0$ otherwise. Here, $\diag(\bm a)$ denotes a matrix with diagonal given by $\bm a$ and that is zero otherwise. In addition, 
parameter $\tilde{\zeta}_t$ controls the spatial autocorrelation structure: $\tilde{\zeta}_t = 1$ corresponds to the intrinsic CAR prior \citep{Besag1991}, where the conditional expectation is the mean of the random effects in geographically adjacent areal units. On the other hand, $\tilde{\zeta}_t = 0$ corresponds to independent random effects. The class of CAR models is large; see further details in \cite{besag1974spatial}, \cite{cressie1993statistics}, \cite{kaiser2000construction}, \cite{cressie2015statistics}, \cite{riebleretal:16}, and references therein.

We introduce next the concept of \emph{regime}. Given the nature of the data, we do not actually expect the parameters of the model to vary at each time $t$. Instead we expect to observe different states (called regimes), each with a specific set of parameters. A similar notion has been employed in \cite{berliner-wikle-cressie:99}, who consider switching weather regimes when modelling Tropical Pacific sea surface temperatures. The choice of number of regimes in our context is based on information about the evolving dynamics of the system. In our specific application, we assume the number of regimes to be based on the days of the week (weekday/weekend), and on the period of the day (night/day). The regime indicators are denoted by $r_t \in \{1, \dots, n_R\}$, for $t = 1, \dots, T$, where $n_R$ is the number of regimes allowed in the model. In light of the previous discussion, we assume some of the model parameters to be regime-specific, namely the harmonic regression coefficients: at each location
$i$, we have $ \bm \beta_{i{1}}, \ldots \bm \beta_{i {n_R}}$. In addition, we consider $n_R$ $I$-dimensional vectors of spatial random effects, $\bm u_1, \ldots \bm u_{n_R}$, as well as regime-specific scaling parameters $\tau^2_{1}, \dots, \tau^2_{n_R}$  and observation variances $\sigma^2_{\epsilon_1}, \dots, \sigma^2_{\epsilon_{n_R}}$. Finally, the parameters of the spatial precision matrix $Q$ are also regarded as regime-specific, $\zeta_{1}, \ldots, \zeta_{n_R}$. Using the correspondence between time point $t$ and the regime present at that time $r_t$, we have that, at each $t$, $\tilde{\bm \beta}_{it} = \bm \beta_{ir_t}$, $ \tilde{\bm u}_t = \bm u_{r_t}$ and $\tilde{\zeta}_t = \zeta_{r_t}$. This implies a substantial reduction in the number of model parameters. In other words, we distinguish between $\{\tilde{\zeta_t}\}$, the time-specific spatial effects, and 
$\zeta_{r_t}$, the regime-specific spatial association parameter, while establishing a simple mapping among these. A similar argument applies to the time-dependent regression parameters and the random effect parameters.
Conditionally on regimes, we can rewrite the model as follows:
\begin{align*}
&Y_{it} \mid \bm x_t, \bm \beta_{i r_t}, u_{ir_t} = \bm x'_t \bm \beta_{ir_t} + u_{ir_t} + \epsilon_{it},  \quad
\epsilon_{it}  \ind \text{N}\left(0, \sigma^2_{\epsilon_{r_t}}\right)\\
&\bm u_{r_t}\mid \tau^2_{r_t}, Q\left(\zeta_{r_t}, W\right), \zeta_{r_t} \sim \text{N}_I\left(\bm 0,  \tau^2_{r_t} Q\left(\zeta_{r_t}, W\right)^{-1}\right)
\end{align*}
where each $\bm \beta_{i r_t} $ is $p$-dimensional, while $\bm u_{r_t} $ is an $I$-dimensional parameter vector. The marginal priors for the parameters, as well as the whole model specification, are given in Section~\ref{sec:st_clustering}.

\subsection{Areal Product Partition Model (aPPM)}
\label{sec:aPPM}

We next discuss the time-varying spatial clustering structure component of
the proposed model for the $I$ time series of responses across regions. As commonly done in Bayesian nonparametrics, we define a clustering model
by considering a prior distribution for the random partition parameter $\rho_r = \{C_1^r, C_2^r, \ldots, C_{K_r}^r\}$ that denotes the partition of areas $\{1,2,\ldots,I\}$ in the sample at regime $r$. We define a prior distribution by building on product partition models~\citep[PPM, see, e.g.,][]{quintana2003bayesian}, and by modifying the spatially-oriented PPM proposed in~\cite{Hegarty_Barry2008}. To explain the proposal, we recall these models here; for notational convenience, we drop the regime subscript $r$ until the end of this section. Under a PPM, the distribution on a partition $\rho$ of a set of indices $[I]=\{1,\ldots,I\}$ into $K$ subsets $\rho= \{C_1, C_2, \ldots, C_{K}\}$
is constructed in terms of a {\em cohesion} function for any subset $C_j\subset [I]$, which measures the strength of prior belief that the elements of $C_j$ are to be grouped together. The PPM prior is thus expressed as 
\begin{equation*}
p(\rho=\{C_1,\ldots,C_K \}) = \mathcal{K}\prod_{j = 1}^{K} c\left(C_j\right),\qquad \rho\in\mathcal{P}\left([I]\right)
\end{equation*}
where $\mathcal{P}([I])$ denotes the set of all partitions of $[I]$
and $\mathcal{K}$ is an
appropriate normalizing constant depending on the cohesion function. A typical choice of cohesion function is $c(C_j) = \kappa \times \Gamma(n_j)$, with $n_j = |C_j|$, for $j = 1, \dots, K$ from which we recover the exchangeable partition probability function (EPPF) corresponding to the Dirichlet process (DP) with mass parameter $\kappa > 0$. 

Recent developments in the study of PPMs have explored the possibility of adding additional prior information to the definition of $c$, such as covariates \citep{muller2011product, page2016spatial}. In our approach, we include the information about the areal structure of the data, and we do so by following the idea of \cite{Hegarty_Barry2008}, who introduce the notion of \emph{boundary length} $\ell^j (i)$ of the $i$-th areal unit belonging to cluster $C_j$ as the number of neighbours of $i$ that do not belong to $C_j$, for $j = 1, \dots, K$. The boundary length of a cluster $C_j$ is computed as the sum of the boundary length of each areal unit in $C_j$. The model by~\cite{Hegarty_Barry2008}
is thus given by
\begin{equation}
p(\rho=\{C_1,\ldots,C_K \} \mid \xi) = \mathcal{K}\left(\xi, I\right) \prod_{j = 1}^K \e^{-\xi \sum_{i \in C_j}\ell^j (i)}
\label{eq:HegartyBerry}
\end{equation}
where the normalizing constant $\mathcal{K}\left(\xi, I\right)$ is a function of $I$ and of the positive hyperparameter $\xi$, so that
\begin{equation*}
\frac{1}{\mathcal{K}\left(\xi, I\right)} = \sum_{\rho\in\mathcal{P}\left([I]\right)}  \prod_{j = 1}^K \e^{-\xi \sum_{i \in C_j}\ell^j (i)}
\end{equation*}

As a prior for the random partition $\rho$, the proposed model combines the DP-based PPM and the spatial prior by \cite{Hegarty_Barry2008}, with the difference that we consider as neighbours the eight areal units surrounding $i\in[I]$, and not just the ones with a common side. Concretely, we consider
 \begin{equation}\label{eq:aPPM_prior}
 p\left(\rho=\{C_1,\ldots,C_K \} \mid \kappa, \xi\right) = \mathcal{K}\left(\kappa, \xi, I\right) \kappa^{K} \prod_{j = 1}^{K}
 \Gamma\left(n_j\right) \e^{-\xi \sum_{i \in C_j}\ell^j (i)}
 \end{equation}
where $\mathcal{K}\left(\kappa, \xi, I\right)$ is now a function of the positive hyperparameters $\kappa$, $\xi$ and the number $I$ of items to cluster. Compared to \eqref{eq:HegartyBerry}, prior \eqref{eq:aPPM_prior} introduces an additional term, that coincides with that of the product form arising from the Dirichlet process prior when written in PPM form. In particular, following the notation in \cite{Hegarty_Barry2008}, the normalizing constant $\mathcal{K}(\kappa, \xi, I)$ is such that 
\begin{equation*}
\frac{1}{\mathcal{K}\left(\kappa, \xi, I\right)}= \sum_{\rho\in\mathcal{P}\left([I]\right)} \prod_{j = 1}^{K} \kappa\Gamma\left(n_j\right) \e^{-\xi \sum\limits_{i \in C_j}\ell^j (i)}
\end{equation*}
Thus, the number of clusters a priori grows with $\kappa$. On the other hand, $\xi$ is related to spatial association by penalizing excessive spatial disconnectedness between subsets (i.e. small values for the boundary length), in the sense that a larger $\xi$ encourages fewer clusters. The combination of both approaches strikes a balance between a ``rich gets richer'' DP-based clustering and a spatially-oriented setting proposed by \cite{Hegarty_Barry2008}. As shown in Supplementary Section 2, we conduct simulation studies aimed at understanding the intertwining role of $\kappa$ and $\xi$ in controlling the prior partition structure.  

\begin{figure}[ht]
	\centering
	\includegraphics{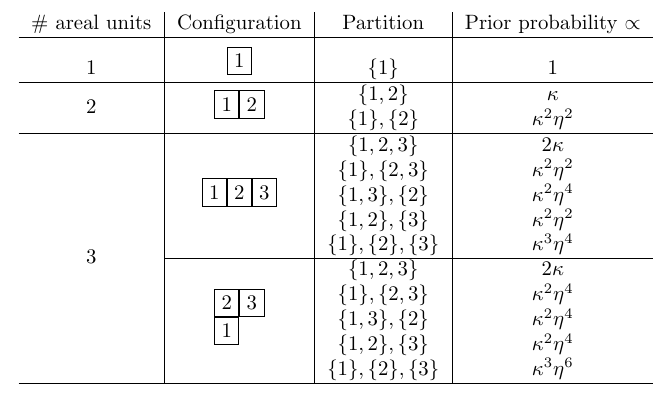}
	\caption{Prior probabilities (up to adequate normalising constants) for different configurations of one, two or three adjacent areal units. The probabilities are expressed as function of the mass parameter $\kappa$ and the boundary length parameter $\eta = \exp(-\xi)$.}\label{tab:PriorArealunits}
\end{figure}

To further understand the partition structure imposed by the proposed model \eqref{eq:aPPM_prior}, set $\eta = \exp(-\xi) \in (0,1)$ and consider one, two or three areas, i.e., $I=1,2,3$. Table~\ref{tab:PriorArealunits} shows the prior probability for different configurations given by the proposed prior distribution, up to a normalizing constant. When $\eta = 1$ (equivalently, $\xi = 0)$, the partition probabilities reduce to the DP prior case. This means that, a-priori, $\kappa > 1$ will assign more probability to partitions composed of singletons, while $0 < \kappa < 1$ to the partition with only one cluster. Considering a fixed value of $\kappa > 1$, similar considerations can be made for the value of $\eta$. Notice that the partition with one cluster is always favored when $\kappa \leq 1$ and that, in the case of three areas depicted in Table~\ref{tab:PriorArealunits}, two different areal configurations are allowed, yielding different a-priori probabilities. In particular, the partition with three clusters is penalised more when the areal units are not aligned, since in our 8-neighbours setting all units are in contact with each other. Secondly, the partitions with two clusters in the second scenario have the same prior probabilities, while in the first scenario the prior probability depends on the specific partition. The partition in which areas 1 and 3 are clustered together is less probable than the one in which 1 and 2 or 2 and 3 are clustered together. Moreover, when $\kappa$ is fixed and $\eta \to +\infty$, the probability of the partition with one cluster tends to one. This simple study shows the spatial ``local'' effect of the parameter $\eta$ on the clustering.

When the number of areal units is larger than $I=3$, it is difficult to see analytically what this implies. We propose in this section a simulated example aimed at understanding the properties of the distribution of the partition a-priori. We show results applied to a rectangular grid of size 14x13, the same as the one used in the Telecom data application, yielding $I = 182$ areal units. The sensitivity analysis reported below presents the distribution of the number of clusters and the properties of the partition a-priori under the proposed model.
For comparison, we also consider the \cite{Hegarty_Barry2008} prior.
MCMC samples from this model follow easily from the Gibbs sampler algorithm
in Section 1 of Supplementary Material by suitably dropping the likelihood and DP parts. For the proposed model, we fix $\kappa$ to 1 so that the expected number of clusters under the regular DP prior is $\mathbb{E}\left(K\right) = \sum_{j = 1}^I \kappa/\left(\kappa + j - 1\right) \approx 5.78$. We also select a grid of values for $\xi$ for comparison; see Supplementary Figures 13 and 14.

In \cite{Hegarty_Barry2008}, the authors report that their prior distribution induces partitions with fewer large clusters for large values of $\xi$ and vice-versa, small values of $\xi$ induce partitions with many clusters of reduced sizes. In particular, Supplementary Figure 13
shows the prior distribution of the number of clusters for a grid of $\xi$ values for the \cite{Hegarty_Barry2008} prior setting. As we can observe, the distribution of the
number of clusters is concentrated on lower values as the value of $\xi$ increases.

The behaviour of our partition prior is summarized in Supplementary Figure 14 by reporting the prior distribution of the number of clusters. We can observe how the effect of $\xi$, as in the previous simulation, is to produce coarser partitions. However, this effect is incremented by the presence of the DP part in our prior specification, evident by comparing these results with Supplementary Figure 13, where the prior number of clusters is, given the same value of $\xi$, much higher.

\subsection{Regime- switching aPPM - Time varying spatial clustering}
\label{sec:st_clustering}

As mentioned in Section~\ref{sec:intro}, we are interested in detecting which areal units share similar temporal patterns. For this reason, the strategy we  develop below builds the clustering structure on the array of coefficients $\bm \beta = \left[ \bm \beta_{ir_t} \right]$, for $i\in\left[I\right]$, $t\in\left[T\right]$, and associated regime $r_t \in \left[n_R\right]$, in the spirit of Bayesian nonparametric priors of discrete nature. Thanks to the introduction of the concept of  model regimes, the clustering induced on the observations via $\bm \beta$ is able to reflect changes appearing through time, allowing for a regime-specific prior distribution for the partition of areas.

\begin{figure}[ht]
	\centering
	 \includegraphics[width=1\textwidth]{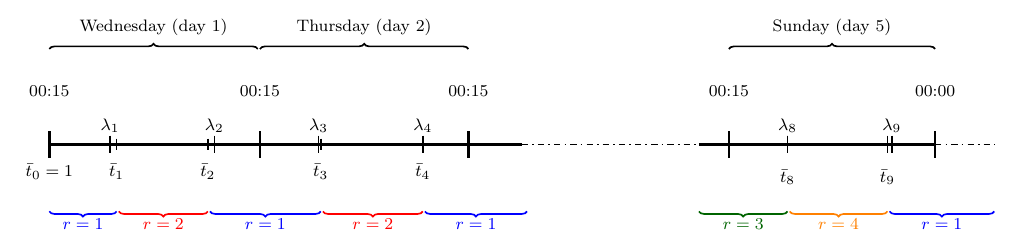}
    \caption{Illustration of the regime scheme in the case under study for the first 5 days. Throughout the time interval, the regimes are depicted in different colors, here indicating night/day separation. Notice how the regime interval is spanned by the points $\bar{t}_r$, which are not necessarily coinciding with the central time points $\lambda_r$. The total number of regimes is 4. 
    }\label{fig:tikz_regimes}
\end{figure}

Firstly, we elaborate on the concept of regime and its relationship with the clustering of observations. A regime $r_t$ at time $t \in \left[T\right]$ is associated to a partition of the areas indexed in $\left[I\right]$ indicated as $\rho_{r_t} = \{C^{r_t}_1, \dots, C^{r_t}_{K_{r_t}}\}$. The number of clusters for each regime-specific partition is denoted by $K_{r_t}$. Because we allow for the same regime to exist at multiple time points over the time period under study, we must permit the same partition $\rho_{r_t}$ to exist at these time points, effectively exploiting the division of the set $\left[T\right]$ of time points into $M \geq n_R$ non-overlapping sets. For instance, in our application, we distinguish between night and day regimes, as well as weekday/weekend, therefore considering $n_R = 4$ distinct regimes. However, we consider $M = 15$ non-overlapping intervals, since we have two weeks of observations, and there are two changes within each day to separate day by night (see Figure~\ref{fig:tikz_regimes}). Of course, this choice is suggested by the specific application considered here, but either simpler or more general choices could be considered. Changes in the regime status are identified by the time points $\bar{t}_0, \dots, \bar{t}_M$, dividing  the time set into non-overlapping intervals, such that:
\begin{equation}
\left[T\right] = \{ \bar{t}_0,2,\ldots, \bar{t}_1\} \cup \{\bar{t}_1+1,\ldots,\bar{t}_2 \} \cup \cdots \cup \{\bar{t}_{M-1}+1,\ldots,\bar{t}_M \}
\label{eq:timepartition}
\end{equation}
where $\bar{t}_0 = 1$, $\bar{t}_M = T$ and $\bar{t}_1< \bar{t}_2<\cdots <\bar{t}_{M-1}<\bar{t}_M$. Depending on the application under study, one might possess prior information on when regimes change. However, in many situations such as the one presented here, the time points $\bar{t}_1, \dots, \bar{t}_{M-1}$ are not deterministically known. Hence, we impose a prior distribution over a range of possible time points 
within each day of the week. Recalling that $\bar{t}_0 = 1$ and $\bar{t}_M = T$, the centre of the regime change intervals are denoted by $\lambda_m$, for $m = 1, \dots M-1$. Specifically, we assume a discrete uniform prior distribution over  $2n_{\lambda} + 1$ discrete time points. 
The regime change interval centres can be fixed according to the prior belief or estimated 
from the data themselves.
For example, if $\lambda_1$ has been fixed or estimated to be 7:00am, we could set $n_{\lambda} = 4$ and assume $\bar{t}_1$, the first point of regime changes, to be uniformly distributed over the time points corresponding to 6am, 6:15am, 6:30am, 6:45am, 7am, 7:15am, 7:30am, 7:45am, and 8am, representing the prior belief of the city awakening period. 
Of course, we always assume \eqref{eq:timepartition} to be a partition of $\left[T\right]$. This is imposed by selecting an integer $n_{\lambda}$ such that the support of each variable $\bar{t}_m$ is $\{\lambda_m - n_{\lambda}, \dots, \lambda_m + n_{\lambda}\}$. Notice that, given the values $\bar{t}_m$, for $m = 0, \dots, M$, the regime status $r_t$ at each time point is known deterministically. 

Finally, we assume that, conditionally on all the change-of-regimes points $\bar{\bm t} = (\bar{t}_0, \dots, \bar{t}_M)$, the $n_R$ regime-specific partitions $\rho_1,\ldots,\rho_{n_R}$ are independent, each distributed according to \eqref{eq:aPPM_prior}. Each regime-specific partition $\rho_r = \{C^r_1, \dots, C^r_{K_r}\}$, for $r \in \left[n_R\right]$ can be equivalently represented through the introduction of a regime-specific vector of allocation variables $\bm s^r = (s^r_1,...,s^r_I)$. Furthermore, a vector of unique values $\bm \beta^*_r = (\bm \beta^*_{1r}, \dots, \bm \beta^*_{K_rr})$ for each regime can be linked to $\tilde{\bm \beta}$, and in such a way to the observations, so that $s^{r}_i = j \iff \tilde{\bm \beta}_{it} = \bm \beta^*_{j{r}} \iff i \in C^{r}_j$ and $r = r_t$. 

After introducing the allocation vector $\bm s^r = (s^r_1,...,s^r_I)$, and the set of unique parameter values $\bm \beta^*_r$, we can describe the predictive distribution of $\bm s^r$. For notational convenience, we temporarily omit the index $r$ indicating the regime. Consider the partition $\rho^i = \{C^i_1, \dots, C^i_{K^i}\}$ obtained by clustering the first $i$ elements into $K^i$ clusters. The predictive law of a new element $i+1$ can be computed as follows:
\begin{align}
\label{eq:pred_clustalloc}
&P\left(s_{i+1} = j\mid s_1, \dots , s_i\right) = \\
&\left\{
\begin{array}{ll}
\frac{p(\rho^{i+1} = \{C^i_1, \dots, C^i_j \cup \{i+1\}, \dots, C^i_{K^i}\} | \kappa, \xi)}{p(\rho^i = \{C^i_1, \dots, C^i_{K^i}\} | \kappa, \xi)} \propto n^i_j \e^{-\xi \ell^j (\{i+1\})}, & j = 1, \dots, K^i \\
\frac{p(\rho^{i+1} = \{C^i_1, \dots, C^i_{K^i}, \{i+1\}\} | \kappa, \xi)}{p(\rho^i = \{C^i_1, \dots, C^i_{K^i}\} | \kappa, \xi)} \propto \kappa \e^{-\xi \ell^j (\{i+1\})}, & j = K^i + 1
\end{array}
\right. \nonumber
\end{align}
where $n^i_j = |C^i_j|$ is the $j$-th cluster size before we assign the $(i+1)$-th observation, and $K^i$ is the number of clusters identified by $(s_1, \dots, s_i)$ so that $K^i + 1$ identifies the new cluster label. To recover the first line of formula \eqref{eq:pred_clustalloc} above, note that the ratio there is equal to
\begin{multline*}
\frac{\mathcal{K}^r\left(\kappa, \xi, i+1\right)}{\mathcal{K}^r\left(\kappa, \xi, i\right)}
\frac{\Gamma\left(n^i_j+1\right)}{\Gamma\left(n^i_j\right)} \frac{\e^{-\xi \sum_{m\in C^r_j\cup\{i+1\}}\ell^j (m) }}{\e^{-\xi \sum_{m\in C^r_j}\ell^j (m) }} \\
=\frac{\mathcal{K}^r(\kappa, \xi, i+1)}{\mathcal{K}^r(\kappa, \xi, i)}  n^i_j \frac{\e^{-\xi\sum_{m\in C^r_j}\ell^j (m) -\xi l^j(\{i+1\}) }}{\e^{-\xi \sum_{m\in C^r_j}\ell^j (m) }}
\end{multline*}
so that the ratio $\mathcal{K}^r\left(\kappa, \xi, i+1\right)/\mathcal{K}^r\left(\kappa, \xi, i\right) $ does not depend on $j$. Similar calculations hold for the second line in \eqref{eq:pred_clustalloc}.

Finally, we impose the following prior distribution for the vector of
unique values $\bm \beta^*_r$, for $r = 1, \dots, n_R$:
\begin{align*}
&\bm \beta^*_{1r}, \dots, \bm \beta^*_{K_rr} \mid \rho_r, \bm \mu_{\bm \beta_r}, \bm \Sigma_{\bm \beta_r} \iid  \text{N}_{p}(\bm \mu_{\bm \beta_r}, \bm \Sigma_{\bm \beta_r}) \\
&\bm \mu_{\bm \beta_r}, \bm \Sigma_{\bm \beta_r} \sim \text{N}_{p}(\bm \mu_{\bm \beta_r}\mid \bm m_{\bm \beta_r}
\bm \Sigma_{\bm \beta_r}) \prod_{j = 1}^p \text{inv-Gamma}(\sigma^2_{\bm \beta_rj} \mid a_{\bm \Sigma_{\bm \beta_r}}, b_{\bm \Sigma_{\bm \beta_r}})
\end{align*}
where $\bm \Sigma_{\bm \beta_r} = \diag\left(\sigma^2_{\bm \beta_r1}, \dots, \sigma^2_{\bm \beta_r p}\right)$, $r = 1, \dots, n_R$. Here $\text{inv-Gamma}\left(\cdot\mid a,b\right)$ denotes the inverse gamma density with mean $b/(a-1)$. We note that the unique values of the array $\bm \beta^*_r$, associated with the $r$-th regime, are shared by all those coefficients $\tilde{\bm \beta}_{it}$ for which regime $r$ is active, hence for all $t \in \{\bar{t}_m + 1, \dots, \bar{t}_m\}$. We also assume independence among the $\bm\beta^*$s parameters across different regimes.

The final model, which we refer to as regime-switching areal PPM (RS-aPPM), can be described as follows:
\begin{align}\label{eq:lik}
&Y_{1t}, \dots, Y_{It} \mid \bm x_t, \bm \beta^*_{1r}, \dots, \bm \beta^*_{K_rr}, \rho_r = \{C^r_1, \dots, C^r_{K_{r}}\}, \bm s^r, \bm u_r, \sigma^2_{\epsilon_r}, r = r_t \notag\\
&\qquad\qquad\qquad \ind \prod_{j = 1}^{K_{r}}\prod_{i \in C^{r}_j} \text{N}(y_{it} \mid \bm x'_t \bm \beta^*_{s^{r}_ir} + u_{ir}, \sigma^2_{\epsilon_r}), \textrm{ for all } t: r_t = r
\end{align}
\begin{align}
&\mbox{for $r = 1, \dots, n_R$:} \nonumber\\ 
&\bm u_r| \tau^2_r, Q(\zeta_r, W)\sim \text{N}_I(\bm u_r | \bm 0,  \tau^2_r Q(\zeta_r, W)^{-1}) \label{eq:spatialprior_u} \\
&\bm \beta^*_{1r}, \dots, \bm \beta^*_{K_rr} | \rho_r, \bm \mu_{\bm \beta_r}, \bm \Sigma_{\bm \beta_r} \iid \text{N}_{p}(\bm \mu_{\bm \beta_r}, \bm \Sigma_{\bm \beta_r}) \label{eq:priorbeta} \\
&\bm \mu_{\bm \beta_r}, \bm \Sigma_{\bm \beta_r} \ind \text{N}_{p}(\bm\mu_{\bm \beta_r}| \bm m_{\bm \beta_r}, \bm \Sigma_{\bm \beta_r}) \prod_{j = 1}^p \text{inv-Gamma}(\sigma^2_{\bm \beta_rj} | a_{\bm \Sigma_{\bm \beta_r}}, b_{\bm \Sigma_{\bm \beta_r}}) \\
&\qquad\textrm{ with }\bm \Sigma_{\bm \beta_r} = \diag(\sigma^2_{\bm \beta_r1}, \dots, \sigma^2_{\bm \beta_r p}) \\
&\rho_r \ind p(\rho_r | \kappa, \xi, \bar{\bm t}) = \mathcal{K}^{r}(\kappa, \xi, \bm n^{r}) \kappa^{K_{r}}
\prod_{j = 1}^{K_{r}} \Gamma(n^{r}_j)  \e^{-\xi \sum\limits_{i \in C_j}\ell^j (i)} \label{eq:partitionprior} \\
&\mbox{and prior for the regime changes:} \nonumber\\
&\bar{t}_m \ind Unif\{\lambda_m - n_{\lambda}, \dots, \lambda_m + n_{\lambda}\}, \quad m=1, \dots, M-1 \label{eq:bartprior}
\end{align}
with $\{1, \dots, T\} = \{ \bar{t}_0,2,\ldots, \bar{t}_1\} \cup \{\bar{t}_1+1,\ldots,\bar{t}_2 \} \cup \cdots \cup \{\bar{t}_{M-1}+1,\ldots,\bar{t}_M \}$. We complete the prior specification with, for $r = 1, \dots, n_R$: 
\begin{align}
& \tau^2_r \iid \text{inv-Gamma}(a_{\tau^2_r}, b_{\tau^2_r}) \label{eq:tau^2_r}\\
& \sigma^2_{\epsilon_r} \iid \text{inv-Gamma}(a_{\sigma^2_{\epsilon_r}}, b_{\sigma^2_{\epsilon_r}})
\label{eq:sigma^2_eps}
\end{align}
Furthermore, prior independence is assumed among the parameters in the
different equations above. Recall also that $Q\left({\zeta}_r, W\right)= \zeta_r (\diag(W \bm 1) - W) + (1-{\zeta}_r) \mathbb I_I$, where $W$ is the proximity matrix, as previously introduced in Section~\ref{sec:likelihood_spatialeffects}. We could assume $\zeta_r$s random, e.g. beta-distributed. However, it is well-known (see, for instance, \cite{Banerjee_etal_2014}, Section 6.4.3.3, or \cite{goicoa2018spatio}) that this leads to non-identifiability issues. For this reason, in the data application we have fixed ${\zeta}_r=0.95$ for each regime $r$, encouraging spatial association, and have assumed informative marginal priors for $\tau^2_r$  and $\sigma^2_{\epsilon_r}$. See Section~\ref{sec:application-telecom} for the specific choice. 

Let us denote by $\bm \phi$ the vector containing all the model parameters, that is 
$\bm \phi= \left(\bm y^{mis}, \bm \beta, \bm s, \bm \bar{t}, \bm u, \left(\tau^2_1,\ldots,\tau_{n_R} \right), \left(\sigma^2_{\epsilon_1}, \dots, \sigma^2_{\epsilon_{n_R}}\right), \bm \mu_{\bm \beta}, \bm \Sigma_{\bm \beta}\right)$, where 
$\bm s = \left(\bm s^1, \dots, \bm s^{n_R}\right)$ and $\bm \bar{t} =
\left(\bar{t}_0, \dots, \bar{t}_M\right)$. To obtain posterior samples from  the full joint posterior distribution $\pi\left(\bm \phi| \bm y^{obs}\right)$, we implement a sequence of Metropolis-within-Gibbs steps. See Section 1 of Supplementary Material for its full description.  
Note that since there are missing values in the $\log$-Erlang numbers, denoted by $\bm y^{mis}$, we incorporate them in the parameters to be simulated from the full conditionals, i.e. $\bm \phi$ contains $\bm y^{mis}$. See step 1. of the MCMC algorithm in Section 1 of Supplementary Material.  

At first sight, model \eqref{eq:lik}-\eqref{eq:sigma^2_eps} might seem complicated and overly parameterised. However, it is
in reality a \textit{sparse} model since it considers 
only $n_r$ random partition parameters $\rho_1,\ldots, \rho_{n_R}$,
instead of one per each of the $T$ time points. We will see in Section 1 of Supplementary Material that updating
the $n_R$ partitions is computationally expensive, but still
reasonable. However, using $T=1344$ random partitions would have
been computationally prohibitive.

\subsection{Summary of the simulation study}
\label{sec:summary_simulation} 

We perform extensive simulation studies investigating the effect of prior elicitation on the clustering estimation and posterior distributions of the parameters of interest. We provide in this section a summary of these simulations, while more details are reported in Section 2 of the Supplementary Material.

We simulate data from \eqref{eq:lik}-\eqref{eq:spatialprior_u}, that is, we simulate from $\bm x'_t \bm \beta^*_{s^{r}_ir} + u_{ir}$ plus independent and identically distributed errors, varying the 
regression parameter values and the error distribution. The aim of this simulation study is to  evaluate the proposed model through a comprehensive set of simulation experiments designed to assess clustering performance under varying conditions, including model misspecification, missing data, and multiple regimes. For some of the examples, we also compare our cluster estimates with the ones from \cite{Hegarty_Barry2008}, which forgets
the ``rich gets richer'' aspect of the cohesion function \eqref{eq:aPPM_prior}. 

We illustrate some characteristics of our model via simulated data, with one ($n_R=1$) or two ($n_R=2$) regimes. In all scenarios, a grid of size $12 \times 10$ is used ($I = 120$), with varying underlying true regime-specific clustering structure, as well as the error distributions (Gaussian, $t$, or skew normal), allowing for misspecification. 
We also simulate the time-varying covariates, the \textit{true} values of the coefficients $\bm \beta^*$ and of the random effects $\bm u$.
Then, for each areal unit, a time series $\{y_{it},t=1,\ldots,T \}$ of length $T = 100$ is generated. 
In all the simulations, we run the MCMC algorithm described in Supplementary Section 1 for 15,000 iterations, of which the first 13,000 are then discarded as burn-in, and the last 2,000 are thinned to obtain a sample of 1,000 to be used in the posterior inference. 
In all the experiments, the cluster estimate for each regime is computed by minimizing the posterior expectation of the variation of information loss function \citep{meilua2003comparing}. We report the Adjusted Rand Index (ARI) between the true clustering structure and the one estimated by our model for all simulated scenarios; see Supplementary Tables 1-2.  

Across all scenarios, posterior chains for most parameters show stable behaviour, although the variance parameter $\tau^2$ exhibits identifiability issues. Missing data affect clustering accuracy primarily when entire time series are unobserved. In total, we consider eight simulation settings, and their results are summarized next:
\begin{description}
\item[Examples 1-2 (single regime):] assess recovery of spatial clusters under contamination and missing data. The model achieves near-perfect clustering (ARI approximately equal to 1), with slight degradation when data are missing.

\item[Examples 3–4 (two regimes):] evaluate changepoint detection and clustering under correct specification, absence of changepoints, and misspecified prior support. The model accurately recovers changepoints when supported by the prior, but performance deteriorates when changepoints are absent or outside the prior range.

\item[Examples 5–6 (misspecification):] investigate robustness to non-Gaussian errors ($t$ and skew-normal). The model remains effective, though clustering accuracy decreases in heavy-tailed settings.

\item[Examples 7–8:] examine challenging scenarios with weak spatial structure and non-Gaussian noise; the model still successfully recovers clustering and changepoints.
\end{description}

Overall, results show that the proposed approach reliably identifies clustering structures and regime changes, demonstrating robustness to misspecification and missing data, while highlighting sensitivity to prior assumptions on changepoints. 

Finally, Section 3.1 of the Supplementary Material reports cluster estimates for two of the simulated datasets, specifically the ones used in Examples~1 and 6, using some of the distance-based techniques discussed in Section~\ref{sec:intro}. All three alternative methods tested resulted in cluster estimates that do not exactly match the simulation truth. In particular, two methods, MGWR and GWANN, fail to correctly recover the areal structure of the data. The third method, \texttt{ClustGeo}, provides slightly better estimates as it includes spatial information in the distance-based technique.
However, these methods inherit drawbacks typical of distance-based clustering algorithms: the number of clusters needs to be fixed in advance; the choice of dissimilarity matrices and of the linkage distance can strongly affect the results and computational efficiency of the method; absence of uncertainty quantification; and difficulty in handling missing data.

\section{Application to Telecom data}
\label{sec:application-telecom}

In this section we fit our model to the Telecom data described in Section~\ref{sec:data}.  Recall the $\log$-Erlang values were previously standardised. We assume \eqref{eq:lik} - \eqref{eq:spatialprior_u} with $\bm x_t$ given in \eqref{eq:x_harmonics} and we fix $n_R=4$ regimes. Prior specification is as defined in \eqref{eq:priorbeta} - \eqref{eq:partitionprior} and \eqref{eq:tau^2_r} - \eqref{eq:sigma^2_eps}.
We found that the changepoint support centres have a strong influence on the overall inference. Therefore, we adopted an empirical Bayes approach to estimate these centres $\lambda_m$s for $m = 1, \dots, M-1$, via the \texttt{R} package \texttt{ecp} \cite{james2015ecp};
see Section 5 in Supplementary Material for further details.
The estimated values of $\lambda_m$s are in agreement with our prior belief on the changepoint location: for each day, there is typically a first changepoint in the morning due to commuting work/home hours and school or offices opening times, and a similar changepoint in the evening. 
However, our model allows for random changepoints to be estimated from the data.
The diagonal elements of $\bm \Sigma_{\bm \beta_r}$ are a-priori inverse Gamma distributed with mean and variance equal to 1 and 0.01, respectively. The hyperparameters $a_{\tau^2_r}$, $b_{\tau^2_r}$, $a_{\sigma^2_{\epsilon_r}}$, $b_{\sigma^2_{\epsilon_r}}$ in \eqref{eq:tau^2_r} and \eqref{eq:sigma^2_eps} are fixed in order to get informative priors with means and variances equal to 1 and 0.1, respectively. This was done to mitigate possible unidentifiability issues in the posterior estimates of $\tau^2,\sigma_\epsilon^2$ (see, for instance, \cite{Banerjee_etal_2014}, Section 6.4.3.3, or \cite{goicoa2018spatio}). The hyperparameters in the marginal prior for $\rho_r$ in \eqref{eq:partitionprior} are $\xi$ and $\kappa$. From the analytic expression of this prior, and the prior simulation study, we know that the number of clusters increases with $\kappa$, while it decreases with $\xi$. Given the amount of data allocated to each regime, we found that the corresponding  likelihood terms have predominant weight in the marginal posterior distributions of $\rho_r$. Therefore, in an attempt to induce parsimony in the resulting clustering, we conduct the analysis by fixing  $\xi=2$ and $\kappa=1$, inducing a strongly informative prior on $K_r$ as seen in Supplementary Figure 14. See the end of this section where we discuss sensitivity with respect to $\xi$ and $\kappa$ and how to fix them.

\begin{figure}[h!]
	\centering
	\includegraphics[width=0.8\textwidth]{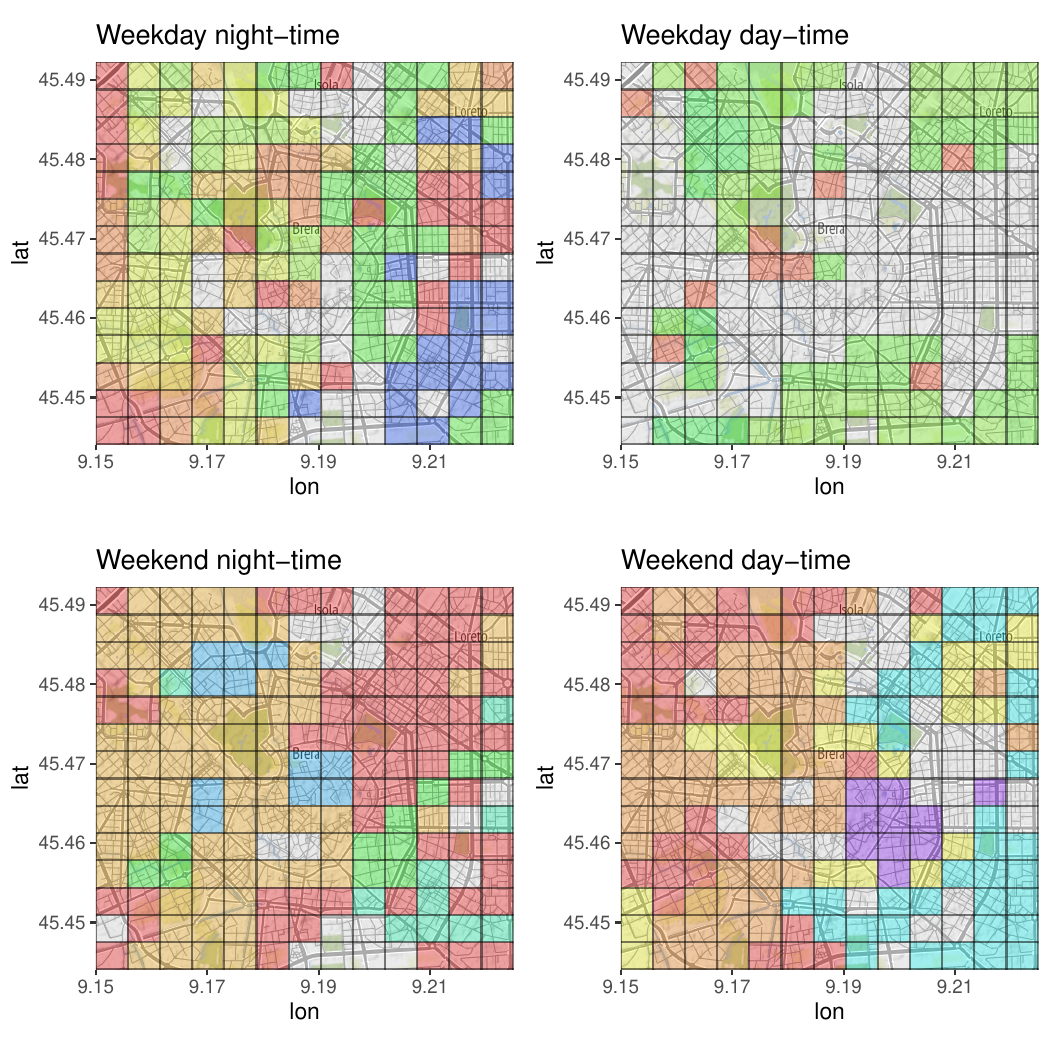}
	\caption{Telecom data. Posterior estimate of the random partition $\rho_r$, for $r = 1, \dots, n_R = 4$, given by minimizing the VI's loss function. Areal units in clusters of size smaller than ten are left uncoloured. Each colour corresponds to a cluster, with the colour scale not reflecting the intensity of a parameter. A map of the total area is superimposed.}
	\label{fig:Telecom_Data_VI_map}
\end{figure}

We run the MCMC for model \eqref{eq:lik}-\eqref{eq:sigma^2_eps} with fixed change-of-regimes times for a total of 50,000 iterations, with a final sample size of 2,500 iterations, after a burn-in of 45,000 and thinning of 2. For each regime, we report a point estimate of the random partition for areal units, minimizing the posterior expectation of the variation of information (VI) loss function with equal misclassification cost parameters. Since the cardinality of the visited partitions is quite large, we employ a hierarchical clustering algorithm with distance equal to the complement of the posterior co-clustering probability and average linkage \cite{Wad18}. Calculations are performed via the R package \texttt{salso} \citep{dahl2022search}; see also \texttt{https://CRAN.R-project.org/package=salso}. 

We show the estimated regime-specific partitions in Figure~\ref{fig:Telecom_Data_VI_map}, where each cluster is identified by a different colour. For visualization purposes, areal units in clusters of size smaller than ten are left uncoloured. We can observe how clusters of large sizes identify regions of the metropolitan area of Milan corresponding to the centre, external rings, and specific hotspots on the map. The grouping of the areal units changes between regimes, reflecting the different types of trajectories observed. Supplementary Figure 15
displays the same plot as in Figure~\ref{fig:Telecom_Data_VI_map}, without the map of the whole area underneath the coloured clustering assignments. 

Figure~\ref{fig:Telecom_Data_VI_map} and Supplementary Figure 15
show some interesting spatio-temporal characteristics, involving major public/travelling spots in the city of Milan. In the regime corresponding to weekend day-time, big clusters are made by contiguous areal units overlapping the more external rings: for instance, 
big clusters are made by contiguous areal units overlapping the more external rings: for instance,
we see a (blue) cluster around Porta Romana and Piazzale Lodi, a portion of the city with many restaurants and bars. There is a large cluster around the area of Porta Venezia, including Piazzale Loreto, which is a shopping
area. The west part of the city is split into two clusters, the largest of them overlapping the more external rings.
For weekend night-time, the city seems to be divided into two larger clusters (the red cluster in the west part and the orange cluster in the east part). During the weekday day-time, it is clear that there is a large part of the city that is split into small clusters (the gray area), but there are two (light and darker green) clusters that split the external rings. Finally, in Figure~\ref{fig:Telecom_Data_VI_map} at weekday night-time, the cluster estimate we get is less homogeneous. For instance, there is a blue cluster at the bottom right corner, overlapping the external ring  (viale Isonzo, piazzale Lodi, viale Umbria), an orange cluster at the top right including Piazzale Loreto.
Note, however, that the total number of estimated clusters is 16 for weekday night-time, 9 for weekend night-time, 12 for weekend day-time and 31  for weekday day-time. For this last regime, comparison to Figure~\ref{fig:Telecom_Data_VI_map} shows that there are many clusters of size smaller than ten (left uncoloured).

In order to understand the differences between the random partitions in the regimes, Table~\ref{tab:posterior_summary_ARI} reports the posterior summary statistics of the Adjusted Rand Index (ARI) between $\rho_i$ and $\rho_j$, $i\neq j$, a similarity measure for partitions, where values closer to 1 indicate partitions that are more similar.
It is clear from the table that the cluster estimates are very different across regimes: in particular, weekday day-time
seems the one less similar to all the others. Table~\ref{tab:posterior_summary_ARI} indicates that incorporating a regime-specific random partition appears to be the most suitable modelling option.

\begin{table}[!ht]
\begin{center}
\begin{tabular}{|*{7}{c|}}
  \hline
  Regimes & 1 Vs 2 & 1 Vs 3 & 1 Vs 4 & 2 Vs 3 & 2 Vs 4 & 3 Vs 4 \\
  \hline
  Quantile $2.5\%$ & 0.100 & 0.162 & 0.134 & 0.026 & 0.063 & 0.363 \\
  Median & 0.104 & 0.169 & 0.137 & 0.029 & 0.064 & 0.369 \\ 
  Quantile $97.5\%$ & 0.108 & 0.176 & 0.140 & 0.030 & 0.067 & 0.375 \\
  \hline
  
\end{tabular}
\end{center}
\caption{Telecom data. Quantiles ($2.5\%$, $50\%$ and $97.5\%$) of the posterior distribution of the Adjusted Rand Index between $\rho_i$ and $\rho_j$, $i\neq j$. Here $i=1$ is weekday night-time, $i=2$ weekday day-time, $i=3$ weekend night-time and $i=4$ weekend day-time.}
\label{tab:posterior_summary_ARI}
\end{table}

\begin{figure}[ht]
	\centering
	\includegraphics[height=0.65\textwidth, width=\textwidth, angle =0]{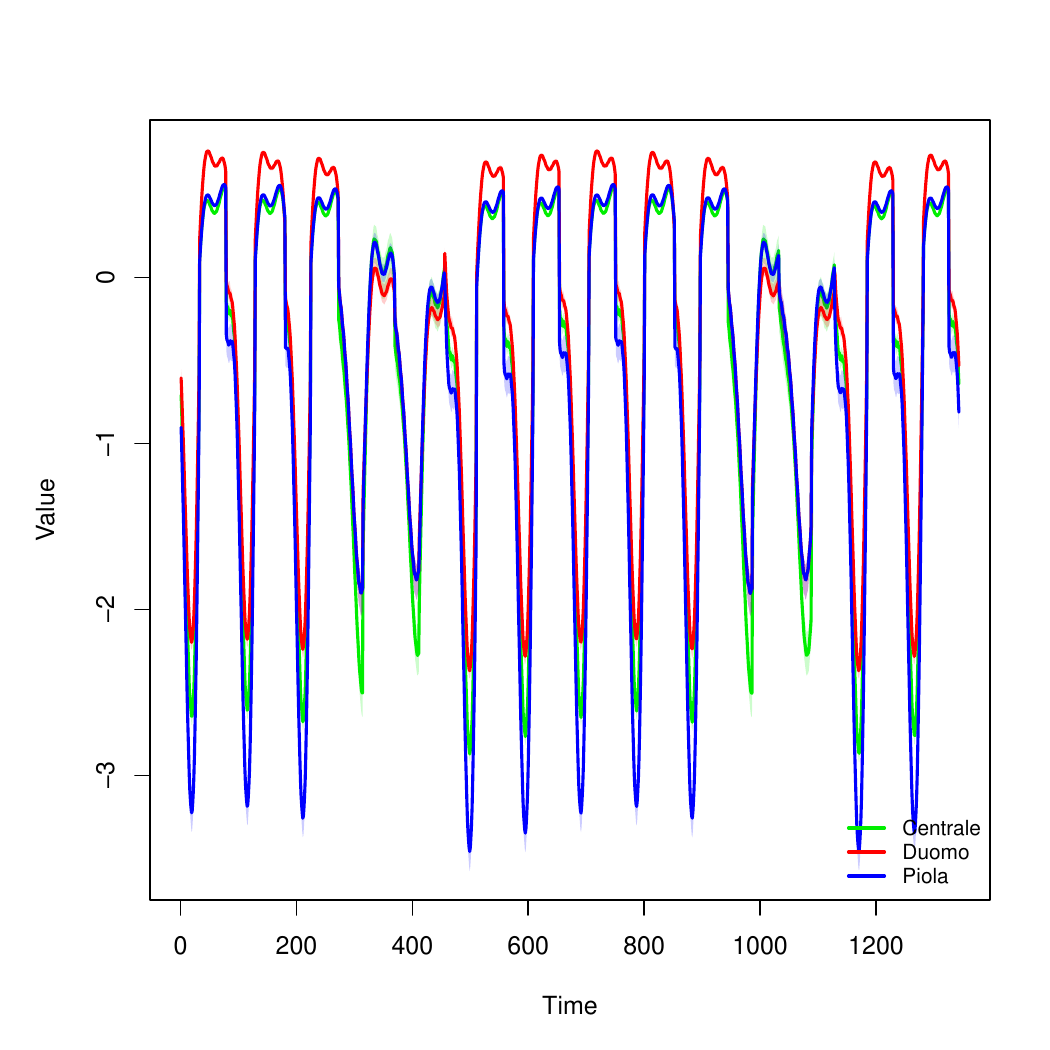}
	\caption{Telecom data. Posterior mean and 95\% credible interval for $\bm x'_t \bm \beta_{ir_t}$, as a function of $t$,   when $i$ denotes the areal unit including Stazione Centrale, Piazza Duomo and Piazzale Piola.}
	\label{fig:xtbeta_selected}
\end{figure}

We display in Figure~\ref{fig:xtbeta_selected} the posterior mean of $\bm x'_t \bm \beta_{ir_t}$ as a function of time $t$, for three different areas corresponding to Stazione Centrale (Central Station), Piazza Duomo and Piazzale Piola, previously shown in Figure~\ref{fig:data_example}(a). These plots suggest that each of these locations exhibit high or low activity, depending on specific periods of the day/week. Duomo is very active on weekdays but less so during weekends, while Piola oscillates between high activity on weekends or low activity late at night.

Section 3.2  of Supplementary Material shows comparison with the  distance-based methods discussed in Section~\ref{sec:intro} for the Telecom data. The Telecom data presents several missing observations which cannot be handled by the alternative methods. Therefore, we impute the missing values via linear interpolation using the R package \texttt{longitudinalData}; see Supplementary Section 3.2 for details on the procedure. However, to overcome issues relative to the amount of memory required to implement these methods, 100 time points were selected uniformly at random from the imputed dataset for further analysis. For one of the methods, GWANN, we only selected ten time points due to memory issues arising with a bigger subset of the data. For one of the methods, \texttt{ClustGeo}, which includes  the spatial information in the distance-based technique, the estimated clusters show some of the structures characterizing
the geography of the metropolitan area of Milan; see Supplementary Figure 11(c). The other two methods 
are able to weakly identify some interpretable features of the partition of areal units,
namely the concentric regions of the map corresponding to the circular structure
of Milan's topology; see Supplementary Figure 12.
As in the case of simulated data, beyond memory issues when using the associated package, these methods need the number of clusters to be fixed in advance and  have no uncertainty quantification.

Section 4 of Supplementary Material shows a performance comparison between the proposed model and some of its variations as well as Bayesian competitors available in the literature. Beyond the full model as described at the beginning of this section, the competing models include the RS-aPPM model equipped with either a Dirichlet Process prior for the partition (full model with $\xi = 0$) or the same prior used in \cite{Hegarty_Barry2008} ($\kappa = 1$ and dropping the $\Gamma(n_j^r)$ terms in \eqref{eq:partitionprior}), and the spatio-temporal conditionally auto-regressive models of \cite{lee2015carbayesst} which are available for implementation through the R package \texttt{CARBayesST}. In particular, for the latter we focus on the \texttt{ST.CARar} model with $\rho = 0, 0.95$. All the models are evaluated on the same Telecom dataset. See Supplementary Table 3 where we report the posterior mode of the number of clusters within
each regime, the LPML and the WAIC for each scenario. In terms of these two indicators, the best model is ST.CARar with $\rho = 0.95$. It is interesting
to observe that the ST.CARar model detects the presence of spatial correlation, as reflected by comparing the cases $\rho=0$ and $\rho = 0.95$. The ST.CARar model includes a more sophisticated and saturated specification of spatio-temporal random effects than our CAR structure, which may explain the LPML and WAIC values here reported.
In terms of goodness of fit, the ranking of the three specifications of the RS-aPPM, after ST.CARar ($\rho=0.95$), is DP version as first, then HB version, followed by the full version (our proposed model) with the lower number of estimated clusters in all regimes, confirming the well-known trade-off between clustering and density estimation. 
Note that all the models from the R package \texttt{CARBayesST}, included ST.CARar, do not allow for any clustering structure estimation, neither do they provide modelling of regimes.

Recall that, a priori, 
$\kappa$ and $\xi$ control the clustering behaviour of our model. This can be seen from the full conditional allocation probabilities of the latent variables $\mathbf{s}^r$ ($r=1,\dots,n_R$), reported in Supplementary Section~1, which are the only quantities whose full conditional involves $\kappa$ and $\xi$. In particular, the probability of creating a new cluster when allocating area $i$ during regime $r$ is proportional to $\kappa e^{-\xi \ell_j(\{i\})}$, where $\ell_j(\{i\})$ is the boundary length of the $i$-th areal unit belonging to the cluster $C_j$, that is the number of neighbours of $i$ that do not belong to $C_j$, for $j = 1, \dots, K_r$. Under our neighbourhood definition, which includes the eight surrounding areas, this probability is bounded below by $\kappa e^{-8\xi}$. Fixing $\xi=2$ therefore keeps the prior probability of creating new clusters small, discouraging excessive fragmentation and preventing the likelihood from dominating the posterior, given the large sample size (about $60,000$ observations per regime). For existing clusters, the allocation probability is proportional to $n_j^{-i} e^{-\xi \ell_j(\{i\})}$, which balances cluster size and spatial cohesion. Our extensive simulations suggest that fixing $\kappa$ at the conventional value of $1$, as often done by default in the BNP literature, the user can obtain better control by moving $\xi$ over its range. We have opted for a conservative prior that favors a smaller number of clusters (as achieved with a relatively large choice of $\xi=2$); see Figure 14 in the SM. Similar results have been obtained with a modest change of these values ($\xi = 1$, results not shown). 
Partitions are similar, though $\xi=1$ gives a larger number of estimated clusters. In the application, with such a large amount of data (approx $60,000$ observations per regime) we had to choose hyperparameters corresponding to a prior concentrated on very small values for each $K_r$. In other applications, if data size were smaller, we could allow a larger prior variability for $K_r$. Summing up, we suggest that the procedure to set $\kappa$
of the same order of magnitude as 1 and fix  
$\xi$ based on the size of the problem and the distribution a-priori of the induced number of clusters.

\section{Summary and Conclusions}\label{sec:conclusions}

Motivated by the analysis of mobile phone usage in part of the city of Milan, Italy, we propose a semi-parametric random partition model for the analysis of large spatio-temporal data. The model features a random partition prior distribution that combines the well-known DP with the HB specifications, providing a balance among spatially cohesive areal grouping and number and sizes of clusters. Due to the nature of the data, consisting of series of measurements every 15 minutes over the entire span of two weeks, the model also incorporates the notion of switching regimes, to
reflect differences over weekday/night and weekend patterns of mobile phone usage. 

Through extensive simulation studies we find that the model fares well when compared to other alternative Bayesian models, including the random partition models corresponding to either the DP or the HB priors (i.e., not combined) and the models included in \texttt{CARBayesST} package available in \texttt{R}. 
We have also compared our model with some non-Bayesian statistical and machine learning approaches which could accommodate or be extended to be used in tandem with spatio-temporal clustering procedures. 
However, none of these methods/packages are able to handle missing data, different temporal regimes, and cluster estimates simultaneously. Moreover, we encountered severe memory issues and we had to drastically reduce the time points (only 100 or even 10 in one case) for the Telecom application.
The overall conclusion is that our approach, being tailored to the particular application at hand, is an all-inclusive approach. Specifically, it is able to account for missing responses, varying regimes corresponding to different times, time evolving responses recorded throughout a number of days, cluster estimates of the areal units, and last but not least, it is capable to quantify uncertainty of the estimates.

We would like to point out that an alternative to doing inference based on
MCMC simulation is to use a Integrated Nested Laplace Approximations (INLA, \cite{RueINLA}). INLA has become increasingly popular in the analysis of spatio-temporal data, also thanks to the R-INLA package. However, while R-INLA can fit many different spatio-temporal models, it does not provide a way to implement a model equipped with a random partition.

A limitation of our model, in fact shared with many related random partition models is the computational burden required to implement posterior simulation via MCMC. However, it has to be noted that the main bottleneck of the code is the number of timestamps, which largely impacts the computational time. We were also able to run our algorithm for about 3 times the number of areas without any dramatic change in the computational time. 
For example, with $182$ areas per $1344$ time points as in Section \ref{sec:application-telecom}, we ran the code on a university server and we recorded an approximate number of 34
iterations/minute. On the other hand, we ran all the simulated examples, consisting of $120$ observation per $100$ time points, on our personal laptops, measuring an approximate 2419
iterations/minute for the two-regimes case.
A possible topic for future research is the adoption of divide and conquer techniques that split the data in smaller segments that can be dealt with in parallel, to be later suitably combined to produce a reasonable approximation to the actual posterior distribution. This approach would yield a faster and more computationally feasible strategy that can be applied to larger datasets, as well as used to fit models featuring a higher number of parameters with the aim of capturing area-specific effects.

\section{Acknowledgments}	
	Andrea Cremaschi acknowledges the Department of Paediatrics, Yong Loo Lin School of Medicine, National University of Singapore and the Singapore Institute for Clinical Sciences (SICS, A*STAR) for the support.  This research was (partially) completed while three authors (Cremaschi, Guglielmi, Quintana) were visiting the Institute for Mathematical Sciences, National University of Singapore, 2024.

\section{Funding}
Andrea Cremaschi acknowledges partial support by PID2024-155187OB-I00 granted by MCIU, and by RYC2024-050330-I, funded by MICIU/AEI/10.13039/501100011033 and FSE+.
Fernando Quintana acknowledges partial support by the grant ANID Fondecyt Regular 1220017. Alessandra Guglielmi has been partially supported by MUR, PRIN project 2022CLTYP4. Giulio Beltramin and Alessandra Guglielmi acknowledge the support by MUR, Grant Dipartimento di Eccellenza 2023–2027.


\bibliographystyle{plainnat}
\bibliography{Bibliography}

\end{document}